\documentclass[%
 aip,
 pof,%
 amsmath,amssymb,
reprint,%
]{revtex4-1}
\usepackage{amsmath,amssymb}
\usepackage{amsfonts}
\usepackage{graphicx}
\usepackage{dcolumn}
\usepackage{bm}
\usepackage{comment}
\graphicspath{{../}}

\begin{document}

\preprint{AIP/123-QED}

\title[Curvature singularity and film-skating during drop impact]{Curvature singularity and film-skating during drop impact}

\author{Laurent Duchemin}
\affiliation{IRPHE, CNRS \& Aix-Marseille Universit\'e, 49 rue Joliot-Curie, 13013 Marseille, France}
\author{Christophe Josserand}
\affiliation{Institut D'Alembert, CNRS \& UPMC (Univ. Paris 06), UMR 7190, Case 162,  4 place Jussieu, F-75005 Paris, France.}

\date{\today }
\begin{abstract}
\textbf{Abstract:} We study the influence of the surrounding gas in the dynamics of drop impact on a smooth surface. We use an axisymmetric 3D model for which both the gas and the liquid are incompressible; lubrication regime applies for the gas film dynamics and the liquid viscosity is neglected. In the absence of surface tension a finite time singularity whose properties are analysed is formed and 
the liquid touches the solid on a circle. 
When surface tension is taken into account, a thin jet emerges 
from the zone of impact, skating above a thin gas layer. The thickness of the air film 
underneath this jet is always smaller than the mean free path in the gas suggesting that the liquid film eventually wets the surface. We finally suggest an aerodynamical instability mechanism for the splash.
\end{abstract}

\maketitle

{\it Introduction:}
Drop impact is present in many surface flows and diphasic dynamics. It is crucial to our
understanding of atomization, ink-jet printing or deposition for instance~\cite{Rein93}, as well as for environmental issues such as raindrop impact erosion or aerosol spreading~\cite{coantic80}. 
The general situation involves an almost spherical drop impacting on a dry or wet solid surface. A splash is observed for strong impact conditions, whereas a gentle spreading of the drop is seen otherwise. On a dry substrate as considered further on, the splash is characterized by a corolla that detaches from the solid substrate forming a corona shape from which droplets can eventually detach~\cite{Rio01}.
This complex dynamics involves various parameters  which can highly influence the transition between these two regimes: primarily viscous and capillary effects are invoked, quantified by the Reynolds and the Weber numbers respectively. The impacted surface
is also important: in particular, the roughness of the surface can control the splash formation~\cite{RF98,Luis05} as enhanced with textured surfaces~\cite{Xu07}.
Recently, the surrounding gas, often neglected in these problems, has been shown to be crucial since the splashing observed at atmospheric pressure disappears when the ambient gas pressure is lowered~\cite{Xu05}. Despite the few models already proposed to explain this striking effect, invoking in particular compressibility of the surrounding gas~\cite{Xu05} or the entrapment of a (compressible) gas bubble by air cushioning~\cite{Mandre09,HiPu10}, a complete understanding for the mechanism of the splash formation is still lacking.
The aim of this paper is thus to disentangle the role of the surrounding gas for the drop impact on solid surfaces in the limits where the viscosity of the liquid of the drop can be neglected and where both fluid can be considered incompressible. 

\begin{figure}
\begin{center}
\includegraphics[width=0.8\linewidth]{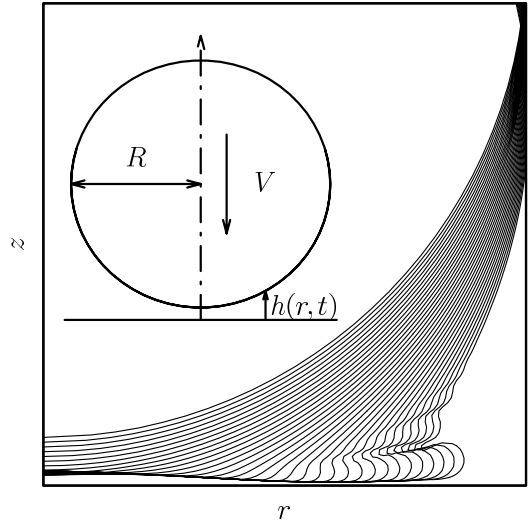}
\end{center}
\caption{\it{Sketch of the impacting drop.}} \label{scalings}
\end{figure}

{\it Governing equations and numerical method.}
We consider the impact of an incompressible liquid drop of radius $R$, density $\rho_l$, with vertical velocity $V$ on a solid substrate (see figure \ref{scalings}). The gas is taken incompressible of density $\rho_g$, dynamic viscosity $\eta$ with surface tension $\gamma$. Although in the experiments \cite{Xu05} the gas is clearly in a compressible regime beneath the drop due to the large pressures created by the air cushioning, we argue that the gas compressibility might not be crucial to understand the splashing transition since it does not change the general structure of the equations.
Gravity can be neglected since the Froude numbers ($Fr=V^2/gR$) are always above $10^{2}$ and axisymmetric approximation can be safely assumed at short time. We note $t=0$ the time at which the drop would touch the wall in the absence of the surrounding gas.
The dominant effect of the gas lies in the dynamics of the viscous thin gas film underneath the drop. In this situation, the viscous boundary layer in the liquid created by the gas shear flow is small and the liquid velocity can be approximated by a potential flow while the dynamics of the gas layer follows the lubrication equation\cite{Korob08,Mandre09}. Our model is thus similar to the 2D approach addressed recently~\cite{Korob08,Mandre09,Mani10}, where a liquid parabola impacts a solid surface, but with crucial differences: 1) it is  3D-axisymmetric; 2) a spherical drop impacts; 3) we use a curvilinear description of the interface which allows for the description of the jet.

Finally, the liquid drop and the gas film dynamics are given by the following dimensionless set of equations in the cylindrical coordinates $(r,z)$:
\begin{eqnarray}
(\partial \Omega) & \partial_t \varphi + \displaystyle \frac12 \nabla \varphi^2 +  p + \frac{1}{{\rm We}} \kappa= C(t), \label{eq:dimbern}\\
(\partial \Omega) &  \partial_t h = \displaystyle \frac{1}{12 r {\rm St}}\partial_r ( r h^3 \partial_r p ), \label{eq:dimlub}\\
(\partial \Omega) & \partial_t h = \partial_z \varphi  -\partial_r \varphi \partial_r h, \label{eq:diminter}\\
(\Omega) & \Delta \varphi= 0, \label{eq:dimincom}
\end{eqnarray}
where $\varphi$ is the velocity potential in the drop (${\bf u}(r,z,t)= {\bf \nabla} \varphi(r,z,t)$), $\Omega$ is the liquid domain, $\partial \Omega$ is the boundary of the drop and $p$ is the lubrication pressure in the gas. The lengths have been rescaled by $R$, velocities by $V$, densities by $\rho_l$ and the gas pressure by $\rho_l V^2$. The full interface $\{r(s,t),z(s,t)\}$ is indexed by the curvilinear coordinate $s$ and $\kappa$ is the mean curvature of the interface. 
This set of equations introduces the two dimensionless numbers of the problem, the Weber and Stokes numbers~:
$$ {\rm We}= \frac{\rho_l R V^2}{\gamma} \;\;\;{\rm and}\;\;\; {\rm St} = \frac{\eta}{\rho_l V R} .$$
Remarkably, general experimental conditions correspond to ${\rm St} \ll 1$ as considered further on.
The incompressibility condition (\ref{eq:dimincom}) joint with the Bernoulli equation at the drop interface (\ref{eq:dimbern})  describes the liquid potential flow. This dynamics is coupled with the surrounding gas flow through the interface advection equation (\ref{eq:diminter}) and the pressure, given in the thin film by the lubrication approximation (\ref{eq:dimlub}). Notice that the lubrication is only valid when the air film is thin enough, otherwise a free surface flow condition ($p=P_0$ constant pressure) has to be applied in the Bernoulli equation. 
The numerical method proceeds as follows~:  Laplace's equation (\ref{eq:dimincom}) is solved using a boundary integral method, the pressure is calculated through the lubrication equation (\ref{eq:dimlub})~\cite{LisDuc08}, the interface and the velocity potential are advanced in time using the kinematic condition (\ref{eq:diminter}) and Bernoulli equation (\ref{eq:dimbern}) respectively. 

\begin{figure}
\begin{center}
\includegraphics[width=\linewidth]{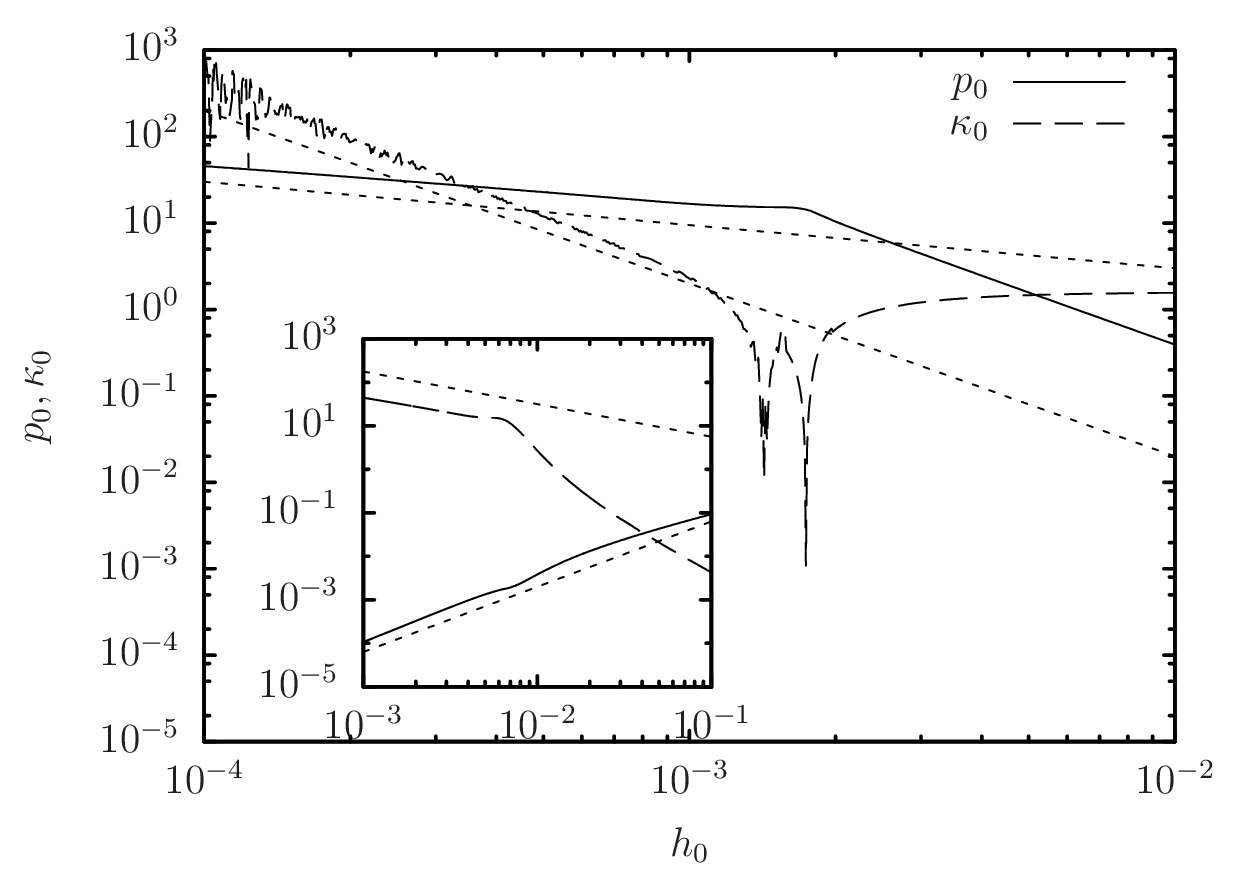}
\end{center}
\caption{\it{$p_0$ and $\kappa_0$ as functions of $h_0$. The two dotted lines show respectively the $h_0^{-\frac12}$ and $h_0^{-2}$ scalings. The inset shows $h_0(t)$ and $p_0(t)$ as function of $(t_0-t)$ close to the singularity, with the fitted scalings $h_0(t)\propto (t_0-t)^{3/2}$ and $p_0(t)\propto (t_0-t)^{-3/4}$.}} \label{fig:sing-prop}
\end{figure}

A sequence of snapshots of the drop impact is shown on figure \ref{scalings} for ${\rm We}=23.7$ and ${\rm St}=1.35 \times 10^{-3}$.
The drop deforms as it approaches the wall and a dimple appears underneath. Then a quasi-horizontal liquid jet expands rapidly. We observe on figure \ref{scalings} that the liquid never touches the wall and the jet "skates" on a thin gas layer! This is consistent with the general property of viscous film that cannot break-up in a finite-time~\cite{Eggers97}. However, this is not the case in the absence of surface tension where corner like interface can be created as described below.\\
\begin{figure}
\begin{center}
\includegraphics[width=\linewidth]{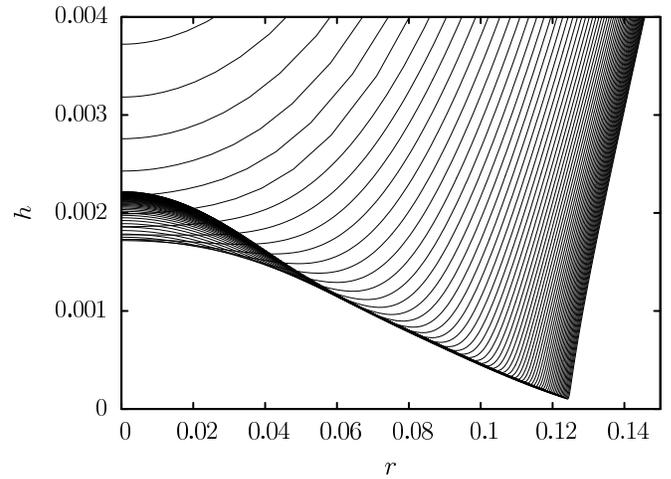}
\end{center}
\caption{\it{Interface profile near the singularity for a drop impact with zero surface tension and ${\rm St}=1.35 \times 10^{-3}$.}} \label{fig:snapsing}
\end{figure}
{\it Finite time singularity for ${\rm We}= \infty$:} as observed in 2D~\cite{Korob08,Mandre09,Mani10}, the dynamics exhibits a finite time singularity in the zero surface tension case, as shown on figure (\ref{fig:snapsing}). It corresponds as $t\rightarrow t_0$ to a corner like interface located at $r_c(t) \rightarrow r_0$ where the curvature $\kappa_0(t)$ diverges as the film thickness $h_0(t)$ vanishes. Similarly, the maximum pressure 
$p_0(t)$ is located in $r_p(t) \neq r_c(t)$ and diverges when $t\rightarrow t_0$ 
(and $r_p(t) \rightarrow r_0$), following:
$$ p_0 \propto h_0^{-\frac12\pm 0.05} \;\; \kappa_0 \propto h_0^{-2\pm0.05}, $$
as presented on figure (\ref{fig:sing-prop}). The radial position $r_0(t)$ of $h_0(t)$ follows approximately the geometrical intersection between the undeformed drop and the solid wall ($r_0\sim \sqrt{2t}$ and $\dot{r}_0=1/\sqrt{2t}$). Using the typical gas film thickness $ H^* ={\rm St}^{2/3} $ by balancing the drop inertia and the lubrication pressure~\cite{Mandre09}, one can estimate $t_0 \sim H^*$ and thus $r_0 \sim {\rm St}^{1/3} $ ($\dot{r}_0\sim {\rm St}^{-1/3} $) at the singularity, in good agreement with the numerical results.

To understand the properties of this singularity, we seek self-similar solutions of the form~:
$\tilde{h}(r,t)=h_0(t) H(R)$, $\tilde{p}(r,t) = p_0(t)P(R)$, and 
$\tilde{\varphi}(r,t)=\varphi_0(t) \Phi(R,Z)$, 
where $R=(r-r_0(t))/l(t)$ and $Z=z/l(t)$. 
Near the singularity, the spatial variations are seen numerically slower than $(t_0-t)$ (see caption of figure \ref{fig:sing-prop}), so that the time derivatives shall be replaced by 
$ (h_0 \dot{r}_0/l(t)) \partial_R$ and equations (\ref{eq:dimbern},\ref{eq:dimlub},\ref{eq:diminter},\ref{eq:dimincom}) give at dominant order, after dropping the tilde~:
\begin{eqnarray}
(\partial \Omega) & - \displaystyle \frac{\varphi_0 \dot{r}_0}{l} \partial_R\Phi+ 
\frac12 \frac{\varphi_0^2}{l^2} \nabla \Phi^2 +  p_0 P = C(t), \label{eq:selfbern}\\
(\partial \Omega) &  - \displaystyle  \frac{h_0 \dot{r}_0}{l} H' = \displaystyle \frac{h_0^3 p_0 ( H^3 P')'}{12  \; {\rm St} \; l^2}, \label{eq:selflub}\\
(\partial \Omega) & - \displaystyle  \frac{h_0 \dot{r}_0}{l} H'  = \frac{\varphi_0}{l} ( \partial_Z \Phi  - \frac{h_0}{l}H' \partial_R \Phi  ), \label{eq:selfinter}\\
(\Omega) & \Delta \Phi= 0, \label{eq:selfincom}
\end{eqnarray}
Taking $\dot{r}_0={\rm St}^{-1/3}$, two regimes can be distinguished, depending on the relevant terms in equation \ref{eq:selfinter}.

{\it Regime I~: $h_0 \ll l$:} 
the second term both in equation \ref{eq:selfbern} and in the right hand side of equation 
\ref{eq:selfinter} can be neglected, and balancing all the other terms, we obtain:
$\varphi_0 \sim {\rm St}^{-1/3}h_0 $, $l\sim  {\rm St}^{-2/3}h_0^{3/2}$ which leads to the observed numerical scalings:
$$ p_0 \sim h_0^{-1/2} \;\;{\rm and} \;\; \kappa_0 \sim {\rm St}^{4/3}h_0^{-2}. $$
We thus deduce that this regime corresponds in fact to {\it thick} gas layer $h_0 \gg  St^{4/3}$.

{\it Regime II~: $h_0 \gg l$:} here the dominant terms balance gives:
$ \varphi_0 \sim   {\rm St}^{-5/3}h_0^2 $, $l\sim  {\rm St}^{-4/3}h_0^2$ and 
$$p_0 \sim {\rm St}^{-2/3} \;\;{\rm and} \;\; \kappa_0 \sim {\rm St}^{8/3}h_0^{-3}.$$
Consistently, this regime holds for {\it thin} gas layer $h_0 \ll {\rm St}^{4/3}$ and it has actually never been reached in all the numerical simulations so far\cite{Smith03,Korob08,Mandre09,Mani10}.
This analysis suggests that the self similar behavior observed in the numerics is valid only for large enough $h_0$ and it predicts that another self-similar regime would hold when approaching the surface closer. In particular, it is not clear whether or not a finite time singularity would still exist (as a cusp then). Moreover,  the lubrication hypothesis would not be valid anymore there, since $h_0/l \gg 1$, and full Navier-Stokes equations should be considered in the gas film.

\begin{figure}
\begin{center}
\includegraphics[width=\linewidth]{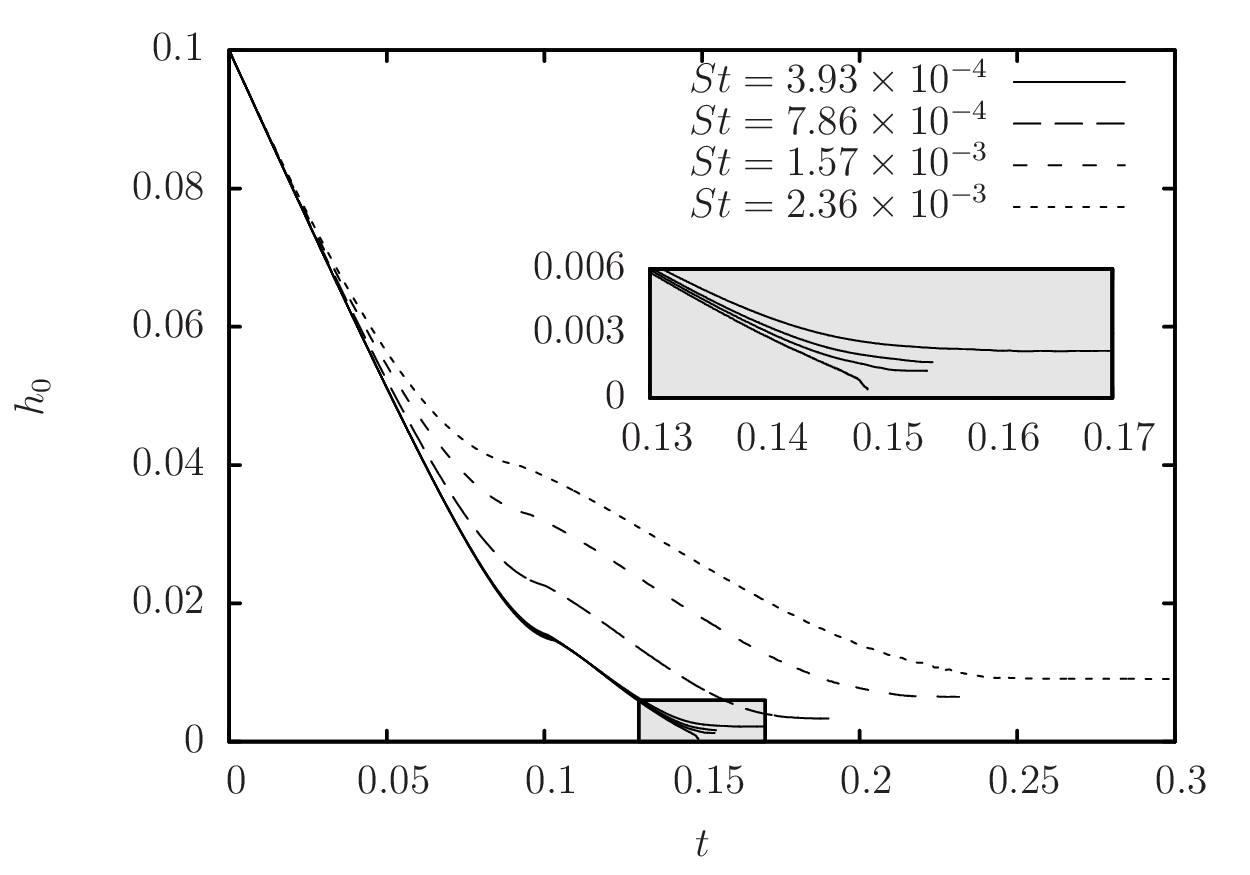}
\end{center}
\caption{\it{Minimum gas film thickness $h_{0}$ 
for four different values of the Stokes number 
($St=3.93\times10^{-4}, 7.86\times10^{-4}, 1.57\times10^{-3}, 
2.36\times10^{-3}$), as a function of time. The zoom for $St=3.93\times10^{-4}$ near the axis shows the evolution for different Weber number, $We=47.5, \,95,\, 238 $ and $\infty$ from top to bottom.}} \label{fig:hmin_t}
\end{figure}

\begin{figure}
\begin{center}
\includegraphics[width=\linewidth]{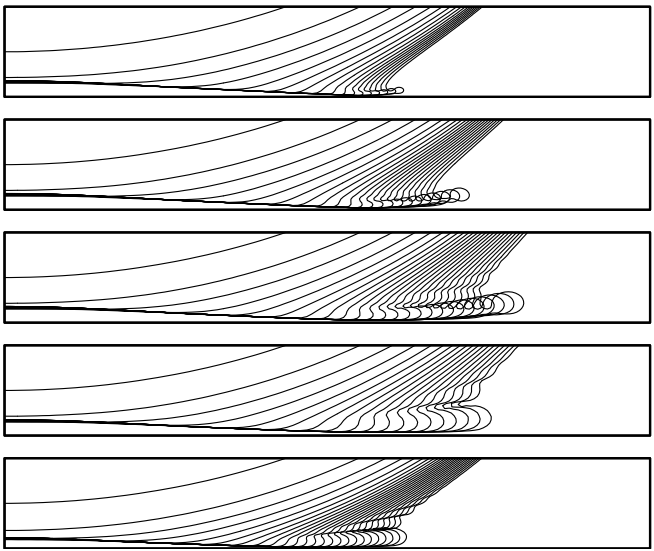}
\end{center}
\caption{\it{Five different computations, with ${\rm We}=238; \; 95; \; 47.5; \; 23.7$ respectively, from top to bottom and ${\rm St}=1.35 \times 10^{-3}$. The bottom figure is for ${\rm We}=23.7$ and ${\rm St}=6.29 \times 10^{-4}$.}} \label{fig:we}
\end{figure}

\begin{figure}
\begin{center}
\includegraphics[width=\linewidth]{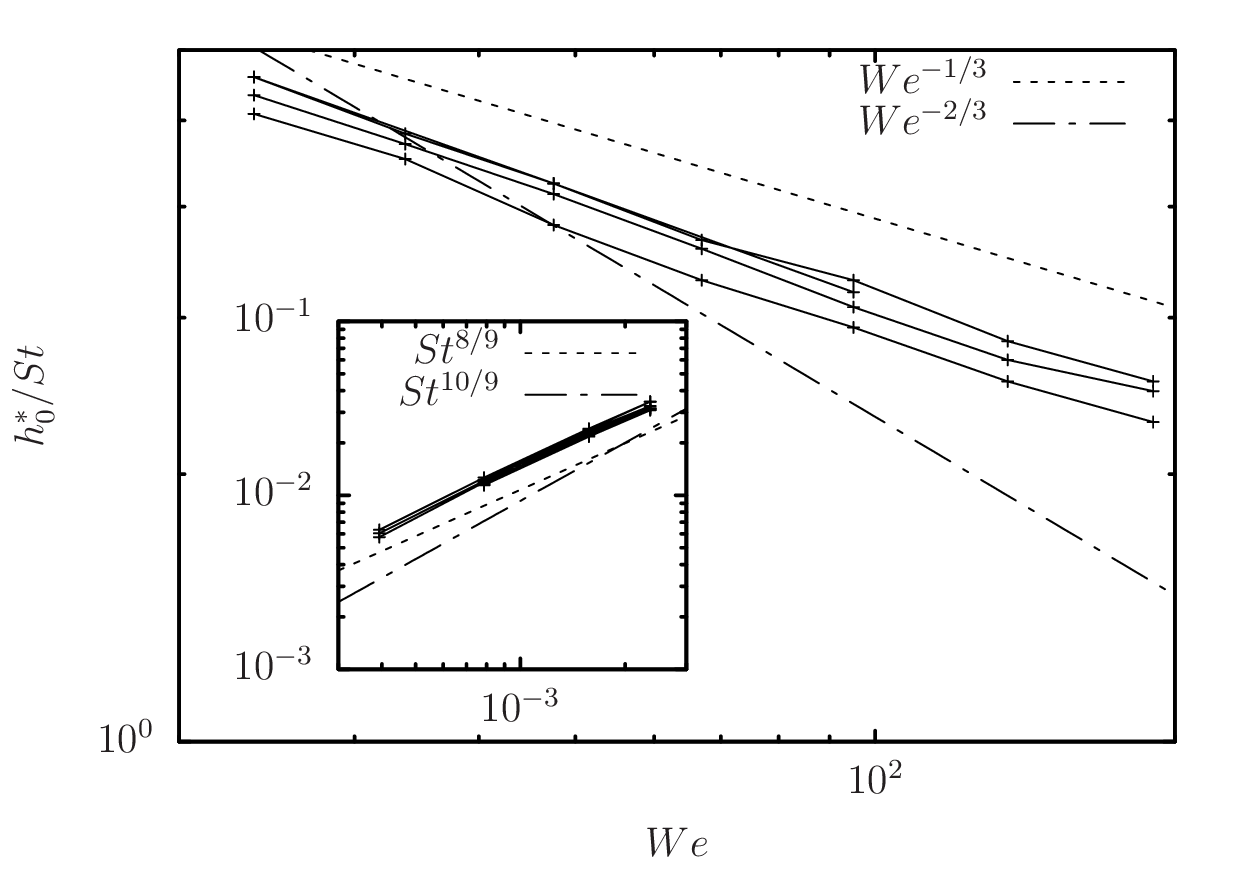}
\end{center}
\caption{\it{Minimum gas film thickness $h_{0}$ 
for four different values of the Stokes number 
($St=3.93\times10^{-4}, 7.86\times10^{-4}, 1.57\times10^{-3}, 
2.36\times10^{-3}$), rescaled by St and 
not-rescaled (inset), as a function of the Weber number.}} \label{fig:hmin}
\end{figure}

{\it Jet formation with surface tension:}
when adding the surface tension, the singularity is regularized since the high curvature regions are smoothed by the capillary
pressure, as illustrated in figure (\ref{fig:hmin_t}) where the minimal air film thickness is shown for different Weber and Stokes numbers. We observe that the dynamics separates from the ${\rm We}=\infty$ case when $h_0$ becomes small enough while the film thickness converges. 
As the Stokes number decreases, the dimple size decreases, and the jet appears earlier with more capillary waves. Moreover, the small angle between the tilted jet and the surface varies both with the Weber and the Stokes numbers (see figure (\ref{fig:we})). The minimal gas layer thickness can be estimated thanks to the two different self-similar regimes exhibited for the singular case ${\rm We}= \infty$. Indeed,  balancing the capillary term ${\rm We}^{-1} \kappa$ with the singular pressure gives for the gas layer depending on the regime: 
$$ {\rm I:}\; h_0 \sim {\rm St}^{8/9} {\rm We}^{-2/3} \;\; {\rm and} \;\; {\rm II:}\; h_0 \sim {\rm St}^{10/9} {\rm We}^{-1/3} $$  while the two behaviors cross for $ {\rm We} \sim {\rm St}^{-2/3}$ (with high Weber number for regime II). Such dependence is investigated on figure \ref{fig:hmin} where the minimal gas thickness is compared with the predicted scalings. We find that the two exponents vary following $h_0 \sim {\rm St}^{0.9-1.} {\rm We}^{-0.33-0.4}$ which is in reasonable agreement with regime II.
Finally, taking the capillary length due to the drop deceleration 
$ \sqrt{\gamma R H^*/\rho_l V^2} \sim {\rm St}^{1/3}{\rm We}^{-1/2}$ for the jet thickness~\cite{Clanet2004}, mass conservation~\cite{JZ03} gives for the jet velocity $V_{jet} \sim {\rm We}^{1/2} {\rm St}^{-2/3} $, in good qualitative agreement with figure (\ref{fig:we}). 

In conclusion, can we explain now the splashing dependence on the gas pressure observed in the experiments~\cite{Xu05}? Since the two control parameters (${\rm We}$ 
and ${\rm St}$) 
do not vary when the gas pressure changes, one has in fact to introduce another physical mechanism. This was somehow the main argument to invoke gas compressibility to explain the experiments although no clear mechanism is identified yet~\cite{Xu05,Mandre09,Mani10}. Here, we propose a purely incompressible mechanism which would be still relevant when gas compressibility will be taken into account~\cite{Mandre09}. Computing the minimal gas thickness in the experiments \cite{Xu05} following the scaling laws for $h_0$ obtained above,
we observe that it is of the order of a few $\AA$, much below the mean free path even at atmospheric pressure ($\sim 60$ nm). Therefore, it is reasonable to consider that the liquid film touches the solid and the splashing mechanism would then be related to the dynamical instability of a rapid thin liquid film expanding on a solid surface. Beside the contact angle dynamics, we want to emphasize that such rapidly expanding film is subject to an aerodynamical instability similar to the Kelvin-Helmoltz one of a liquid jet in a gas environment. If such instability analysis remains to be done, it is likely that it will be dependent on the gas density, by analogy with the classical results of Squire~\cite{squire53}.

Finally, since the scaling laws for $h_0$ are deduced before the jet formation, it is important to notice that liquid viscosity would not
drastically affect the main conclusions of our calculations although it would change the jet thickness (then of order ${\rm St}^{1/3}{\rm Re}^{-1/2}$ in general much smaller than the capillary length calculated before) and the jet velocity (${\rm Re}^{1/2}$)~\cite{JZ03,Mon09}.


\begin{thebibliography}{0}%

\bibitem{Rein93}
M.~Rein.
\newblock Phenomena of liquid drop impact on solid and liquid surfaces.
\newblock {\em Fluid Dyn. Res.}, 12:61, 1993.


\bibitem{coantic80}
Michel Coantic.
\newblock Mass transfert across the ocean-air interface : small scale
  hydrodynamic and aerodynamic mechanisms.
\newblock {\em PhysicoChemical Hydrodynamics}, 1:249--279, 1980.

\bibitem{Rio01}
R.~Rioboo, M.~Marengo, and C.~Tropea.
\newblock Outcomes from a drop impact on solid surfaces.
\newblock {\em Atomization and Sprays}, 11:155--165, 2001.

\bibitem{RF98}
K.~Range and F.~Feuillebois.
\newblock Influence of surface roughness on liquid drop impact.
\newblock {\em J. Colloid. and Interface Science}, 203:pp 16--30, 1998.

\bibitem{Luis05}
C.~Josserand, L.~Lemoyne, R.~Troeger, and S.~Zaleski.
\newblock Droplet impact on a dry surface: triggering the splash with a small
  obstacle.
\newblock {\em J. Fluid Mech}, 524:47--56, 2005.

\bibitem{Xu07}
L.~Xu, L.~Barcos, and S.R. Nagel.
\newblock Splashing of liquids: Interplay of surface roughness with surrounding
  gas.
\newblock {\em Phys. Rev. E}, 76:066311, 2007.

\bibitem{Xu05}
L.~Xu, W.W. Zhang, and S.R. Nagel.
\newblock Drop splashing on a dry smooth surface.
\newblock {\em Phys. Rev. Lett.}, 94:184505, 2005.

\bibitem{Mandre09}
S. Mandre, M. Mani, and M.~P. Brenner.
\newblock Precursors to splashing of liquid droplets on a solid surface.
\newblock {\em Phys. Rev. Lett.}, 102:134502, 2009.

\bibitem{HiPu10}
P.D. Hicks and R.~Purvis.
\newblock Air cushioning and bubble entrapment in three-dimensional droplet
  impacts.
\newblock {\em J. Fluid Mech.}, 649:135--163, 2010.

\bibitem{Smith03}
F.T. Smith, L.~Li, and G.X. Wu.
\newblock Air cushioning with a lubrication/inviscid balance.
\newblock {\em J. Fluid Mech.}, 482:291--318, 2003.

\bibitem{Korob08}
A.A. Korobkin, A.S. Ellis, and F.T. Smith.
\newblock Trapping of air in impact between a body and shallow water.
\newblock {\em J. Fluid Mech.}, 611:365--394, 2008.

\bibitem{Mani10}
M.~Mani, S.~Mandre, and M.P. Brenner.
\newblock Events before droplet splashing on a solid surface.
\newblock {\em J. Fluid Mech.}, 647:163--185, 2010.

\bibitem{LisDuc08}
J.R. Lister, A.B. Thompson, A.~Perriot, and L.~Duchemin.
\newblock Shape and stability of axisymmetric levitated viscous drops.
\newblock {\em J. Fluid Mech}, 617:167--185, 2008.

\bibitem{Eggers97}
J.~Eggers.
\newblock Nonlinear dynamics and breakup of free-surface flows.
\newblock {\em Rev. Mod. Phys.}, 69:865--929, 1997.

\bibitem{Clanet2004}
C.~Clanet, C.~B\'eguin, D.~Richard, and D.~Qu\'er\'e.
\newblock Maximal deformation of an impacting drop.
\newblock {\em J. Fluid Mech}, 517:199--208, 2004.

\bibitem{JZ03}
C.~Josserand and S.~Zaleski.
\newblock Droplet splashing on a thin liquid film.
\newblock {\em Phys. Fluids}, 15:1650, 2003.

\bibitem{squire53}
H.B. Squire.
\newblock Investigation of the instability of a moving liquid film.
\newblock {\em Br. J. Appl. Phys.}, 4:167--169, 1953.

\bibitem{Mon09}
A.~Mongruel, V.~Daru, F.~Feuillebois, and S.~Tabakova.
\newblock Early post-impact time dynamics of viscous drops onto a solid dry
  surface.
\newblock {\em Phys. Fluids}, 21:032101, 2009.

\end{thebibliography}
\end{document}